\documentclass[conference]{IEEEtran}
\IEEEoverridecommandlockouts

\usepackage{cite}
\usepackage{amsmath,amssymb,amsfonts}
\usepackage{algorithmic}
\usepackage{graphicx}
\usepackage{textcomp}
\usepackage{xcolor}
\def\BibTeX{{\rm B\kern-.05em{\sc i\kern-.025em b}\kern-.08em
    T\kern-.1667em\lower.7ex\hbox{E}\kern-.125emX}}

\usepackage{amsmath,amsfonts,amssymb}
\usepackage{algorithmic}
\usepackage{algorithm}
\usepackage{array}
\usepackage[caption=false,font=normalsize,labelfont=sf,textfont=sf]{subfig}
\usepackage{textcomp}
\usepackage{stfloats}
\usepackage{url}
\usepackage{verbatim}
\usepackage{graphicx}
\usepackage{cite}
\usepackage{wasysym}

\usepackage{physics}
\usepackage{glossaries}
\usepackage{xcolor}

\usepackage{mathtools}   
\usepackage{bm}          

\newacronym{qkd}{QKD}{Quantum Key Distribution}
\newacronym{tfqkd}{TF-QKD}{Twin-field Quantum Key Distribution}
\newacronym{tn}{TN}{Trusted Node}
\newacronym{skl}{SKL}{Secret Key Length}
\newacronym{los}{LOS}{Line of Sight}
\newacronym{gs}{GS}{Ground Station}
\newacronym{isl}{ISL}{Inter-Satellite Link}
\newacronym{leo}{LEO}{Low-Earth-Orbit}
\newacronym{sns}{SNS}{Sending-or-not-sending}

\begin{document}

\title{End-to-End QKD Using LEO Satellite Networks \\

\thanks{Copyright © 2026 IEEE. This material may be used for personal purposes. Any other use, including reproduction or redistribution, requires prior permission from IEEE. Requests for such permission should be sent to pubs-permissions@ieee.org.\par This work has been accepted for presentation at IEEE QCNC 2026. \\ The authors thank the Q-net-Q Project which has received funding from the European Union’s Digital Europe Programme under grant agreement No 101091732, and is co-funded by the German Federal Ministry of Research, Technology and Space (BMFTR). This work was further financed by the DFG via grant NO 1129/2-1, by the state of Bavaria via the 6GQT project and by the BMFTR via grants 16KISQ077, 16KISQ168, 16KISQ039, 16KIS1598K and	16KISQ093. The authors acknowledge the financial support by the Federal Ministry of Research, Technology and Space of Germany in the programme of “Souverän. Digital. Vernetzt.”. Joint project 6G-life, project identification number: 16KISK002.}
}


\author{Sumit Chaudhary, Baqir Kazmi, Janis N\"otzel (Member, IEEE) \\ ~\IEEEmembership{}
        
\IEEEauthorblockA{\textit{Emmy-Noether Group Theoretical Quantum Systems Design} \\
\textit{Technical University of Munich}\\
Munich, Germany \\
\{sumit.chaudhary@tum.de, baqirrkazmi@gmail.com, janis.noetzel@tum.de\} }
}

\maketitle

\begin{abstract}
We propose a satellite-based \gls{qkd} network that enables global-scale, end-to-end secure key exchange without relying on trusted intermediate nodes. The network is formed by a ring constellation of satellites that maintain persistent inter-satellite connectivity and support two configurations: a polar Type-I constellation providing global coverage, and an equatorial Type-II constellation offering continuous, terrestrial-like operation. End-to-end secrecy is achieved through the use of \gls{tfqkd} and a redundant XOR-based key-forwarding protocol, in which each forwarding step incorporates independently generated \gls{qkd} keys from ground–satellite and inter-satellite links. As a result, the final secret key is never exposed to any intermediate satellite, eliminating the single-point vulnerabilities inherent in trusted-node networks.
Scaling the network offers two benefits: improved security and higher key rates. Increasing the constellation size enhances security by forcing an adversary to compromise a larger number of nodes to break the protocol, while simultaneously improving link availability and key throughput. Using realistic uplink and \gls{isl} models, we compute finite-size secret-key lengths based on the \gls{sns}–\gls{tfqkd} protocol. Our results show that the achievable key rates scale favourably with constellation size, with Type-II constellations reaching operational continuity and generating multi-gigabit secret keys per day, demonstrating a practical route toward secure global quantum communication.
\end{abstract}


\section{Introduction}
\gls{qkd} enables two distant parties to establish information-theoretically secure secret keys, with security guaranteed by quantum mechanics~\cite{bennett2014quantum}. As quantum computers increasingly threaten classical cryptographic systems, the mathematically unconditional security offered by \gls{qkd} continues to drive research. However, practical \gls{qkd} is fundamentally limited by the exponential decay of key rates with distance due to channel loss~\cite{Pirandola2017PLOB}. Although quantum repeaters have been proposed to overcome this limitation, the required technologies remain experimentally immature. The introduction of \gls{tfqkd}~\cite{lucamarini2018overcoming} marked a significant breakthrough by surpassing the repeaterless bound and extending secret-key distribution distances to nearly 1000 km, as demonstrated experimentally~\cite{chen2020sending, chen2022quantum, liu2023experimental, pittaluga2021600, wang2022twin}. Nevertheless, even \gls{tfqkd} cannot support global-scale secure communication. Satellite-based \gls{qkd} has therefore emerged as a promising solution, achieving distances beyond 1200 km~\cite{yin2017satellite, liao2017satellite}. However, atmospheric turbulence and limited ground–satellite visibility windows restrict performance and result in intermittent key rates~\cite{Sidhu2022finite, sidhu2023finite}. In the absence of practical long-range alternatives, most satellite networks rely on trusted-node architectures, where compromising a single node can threaten the entire network.

To address these challenges, we propose a satellite-based \gls{qkd} network architecture that enables global coverage, enhanced security, and improved key rates. The design employs a persistent ring constellation ensuring continuous inter-satellite connectivity. At any time, two satellites establish \gls{tfqkd} links with ground stations, while the remaining satellites generate and forward key material. A redundant XOR-based key-forwarding scheme guarantees end-to-end secrecy~\cite{calsi2024end}, ensuring that no intermediate node accesses the final secret key. Security is further reinforced by requiring an adversary to compromise a minimum number of consecutive satellites along each independent path, with this threshold increasing as the constellation scales. Beyond security improvements, the ring architecture enhances link availability and aggregate key rates. Increasing the number of satellites raises the ground–satellite visibility fraction and enables near-continuous connectivity. As the constellation grows, the network can achieve high global secret-key rates, offering a scalable and practical path toward secure quantum communication infrastructure.

Recent works have advanced practical satellite \gls{qkd} systems. Reference~\cite{ecker2021strategies} studies strategies for maximizing secret-key rates under realistic free-space conditions. In~\cite{roger2023real}, a 1~GHz satellite \gls{qkd} system is emulated, demonstrating efficient end-user key retrieval via XOR operations on independently distributed keys. Reference~\cite{li2025microsatellite} reports a microsatellite–portable-ground-station platform integrating essential quantum hardware, advancing scalable constellations. Additionally,~\cite{sidhu2023finite} analyzes key engineering challenges, including finite-size effects and system-level constraints for global-scale networks.

The remainder of this paper is organized as follows. Section~\ref{sec_SOP} describes the standard operating procedure. Section~\ref{sec_Proposed_network} presents the proposed architecture and constellation types. The end-to-end key-forwarding mechanism is detailed in Sec.~\ref{sec_e2e_key_forward}. Section~\ref{sec_network_imp} outlines the implementation, including link-budget modeling and \gls{skl} computation. Performance and security analyses are given in Sec.~\ref{sec_results}, followed by conclusions in Sec.~\ref{sec_discussion}.

\section{Standard Operating Procedures}
\label{sec_SOP}

This section outlines the core operational principles governing the proposed sat-\gls{qkd} network.

\begin{figure}[htbp]
\centering
\includegraphics[width=1.0\columnwidth]{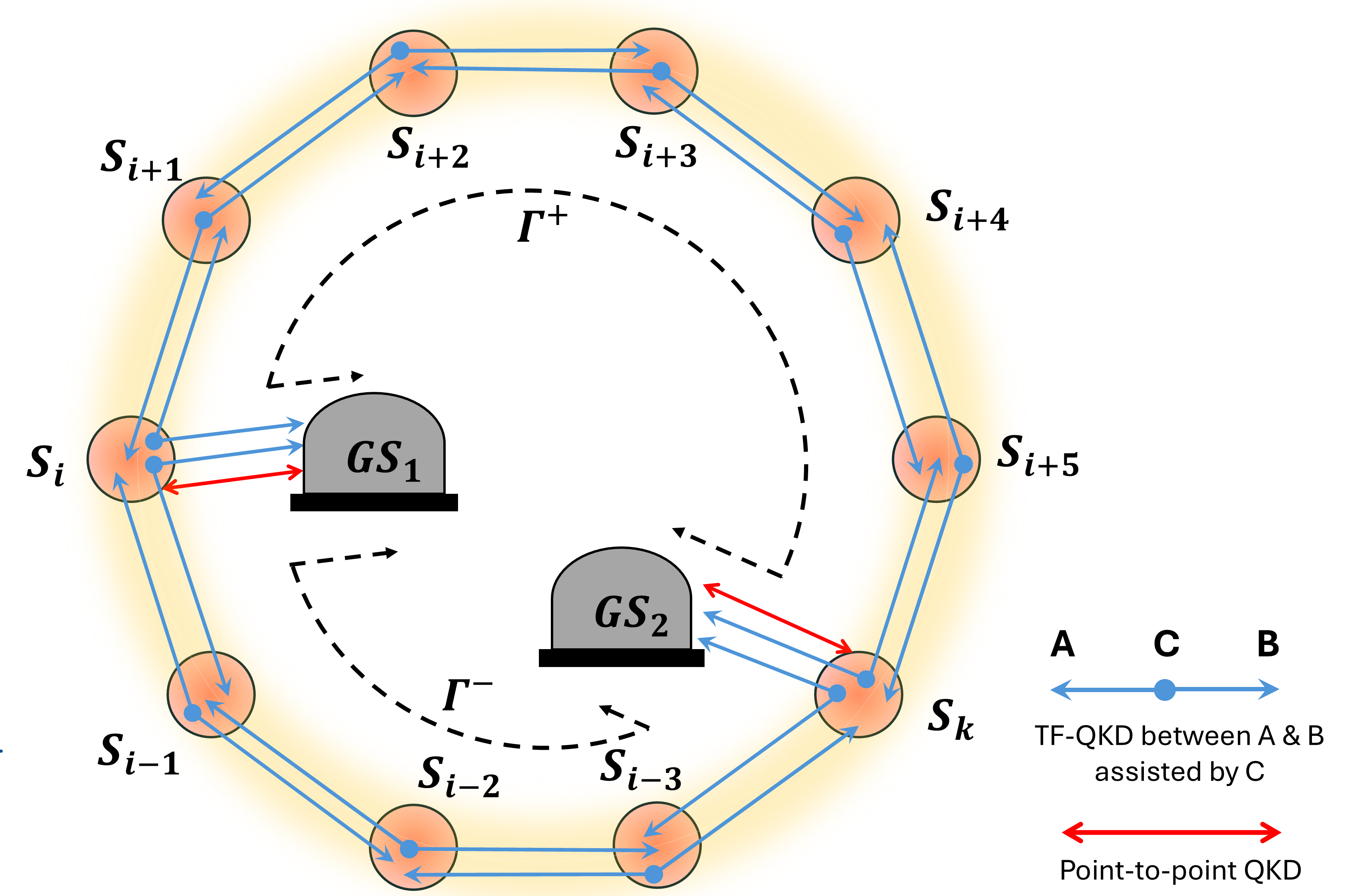}
\caption{Schematics of the satellite-based \gls{qkd} network. Two ground stations, $GS_1$ and $GS_2$, connect with the ring constellation of \gls{leo} satellites. Secret keys $X^+$ and $X^-$ are established along the two directed ring segments, $\Gamma^+$ and $\Gamma^-$, respectively. Blue arrows indicate \gls{tfqkd} links between satellites and between satellites and ground stations, while the red arrow denotes the point-to-point \gls{qkd} links between ground stations and  satellites.}
\label{fig_constellation}
\end{figure}

\textbf{1) Constellation and Hop Structure:}
Satellites are deployed in a ring-type \gls{leo} constellation (Fig.~\ref{fig_constellation}) with uniform inter-satellite spacing. Each \gls{tfqkd} hop forms a tri-nodal configuration consisting of two sending nodes and one central measurement node. The resulting sub-keys are forwarded and combined via XOR operations to establish a multi-hop end-to-end key path.

\textbf{2) Range and Visibility Constraints:}
A \gls{gs} can establish an optical link only when the satellite elevation exceeds $20^\circ$. The number of accessible satellites is therefore limited by \gls{los} visibility constraints, as discussed in Sec.~\ref{sec:III-A-1}.

\textbf{3) Security Breach Threshold:}
Eavesdropping attempts manifest as elevated QBER and result in immediate key discarding. Due to the ring topology and XOR-based forwarding, an adversary must compromise at least two consecutive satellites within each independent path segment to reconstruct the secret key.

\textbf{4) Key Aggregation and Finite-Key Effects:}
The final secret key is obtained by XOR aggregation of sub-keys accumulated over multiple satellite passes. Because \gls{tfqkd} operates under finite-key security bounds, statistically sufficient data must be collected before secure key extraction. In this work, a full key-exchange cycle across the constellation is defined over a 24-hour operational period.

\section{Proposed Network}
\label{sec_Proposed_network}

\subsection{Constellation Model and Link Geometry}

We consider two fixed \glspl{gs} and a ring constellation of $N_s$ identical \gls{leo} satellites 
$\mathcal{S}=\{S_1,\dots,S_{N_s}\}$, where the position of $S_i$ at time $t$ is 
$\mathbf{r}_s^{(i)}(t)\in\mathbb{R}^3$. Cyclic indexing is assumed ($S_{N_s+1}\equiv S_1$).

\subsubsection{Line-of-Sight Constraints}
\label{sec:III-A-1}

\paragraph*{Inter-satellite visibility.}
Adjacent satellites $S_i$ and $S_{i+1}$ maintain unobstructed \gls{los} if the minimum distance between the Earth’s center and the line segment joining them exceeds $R_E+h_{\mathrm{atm}}$. Defining
\[
d_{i,i+1}(t)
=  {\left\lVert \mathbf{r}_s^{(i)}(t)\times \mathbf{r}_s^{(i+1)}(t)\right\rVert} /
{\left\lVert \mathbf{r}_s^{(i)}(t)-\mathbf{r}_s^{(i+1)}(t)\right\rVert},
\]
the condition is $d_{i,i+1}(t) > R_E + h_{\mathrm{atm}}$.

When satisfied for all adjacent pairs, the constellation forms a continuous inter-satellite ring.

\paragraph*{Ground-station visibility.}
Let $GS_j$ ($j\in\{1,2\}$) be located at $\mathbf{r}_g^{(j)}$ with local zenith unit vector $\mathbf{n}_j$. The zenith angle between $S_i$ and $GS_j$ is
\[
\theta_{ij}(t)
=
\arccos\!\left(
\frac{(\mathbf{r}_s^{(i)}(t)-\mathbf{r}_g^{(j)})\cdot \mathbf{n}_j}
{\left\lVert \mathbf{r}_s^{(i)}(t)-\mathbf{r}_g^{(j)}\right\rVert}
\right).
\]
Visibility requires $0 \le \theta_{ij}(t) \le \theta_{\max}$, where $\theta_{\max}$ is the elevation constraint.  
The satellite selected for uplink is

$i'(t,j)=\arg\min_i |\theta_{ij}(t)|$.

A \emph{GS–Sat session} $\tau_k^{(j)}$ is a maximal time interval during which the visibility condition holds continuously for $i'(t,j)$. Its minimum zenith angle is $\bar{\theta}_k^{(j)}=\min_{t\in\tau_k^{(j)}}|\theta_{i'(t,j),j}(t)|$.

The visibility fraction over an observation period $T_{\text{total}}$ is
\[
\rho_{\text{vis}}={\sum_k \tau_k^{(j)}}/{T_{\text{total}}}.
\]

\subsubsection{Satellite Operation and \gls{tfqkd} Functionality}

All satellites support inter-satellite \gls{tfqkd} and satellite–ground \gls{qkd}.  
During a GS–Sat session $\tau_k^{(j)}$, let $S_{i'}$ and $S_{k'}$ be visible to $GS_1$ and $GS_2$, respectively. The network then operates as follows:

\begin{itemize}
    \item $GS_1$ and $GS_2$ perform \gls{tfqkd} with adjacent satellites ($S_{i'\pm1}$ and $S_{k'\pm1}$), assisted by $S_{i'}$ and $S_{k'}$.
    \item All satellites $S_i$ (except $S_{i'}$ and $S_{k'}$) act as central measurement nodes enabling \gls{tfqkd} between $S_{i-1}$ and $S_{i+1}$, maintaining continuous key forwarding around the ring.
    \item Simultaneously, point-to-point \gls{qkd} links are established between each visible satellite and its corresponding ground station.
\end{itemize}

Thus, each satellite concurrently operates as two \gls{tfqkd} senders, two central measurement nodes, and one satellite–ground receiver, ensuring uninterrupted inter-satellite key propagation while opportunistically generating ground keys.

\subsection{Ring Topology}
Now, we describe two types of constellations that form a persistent ring structure with unobstructed line-of-sight at all times.

\begin{figure}[htbp]
\centering
\includegraphics[width=1.0\columnwidth]{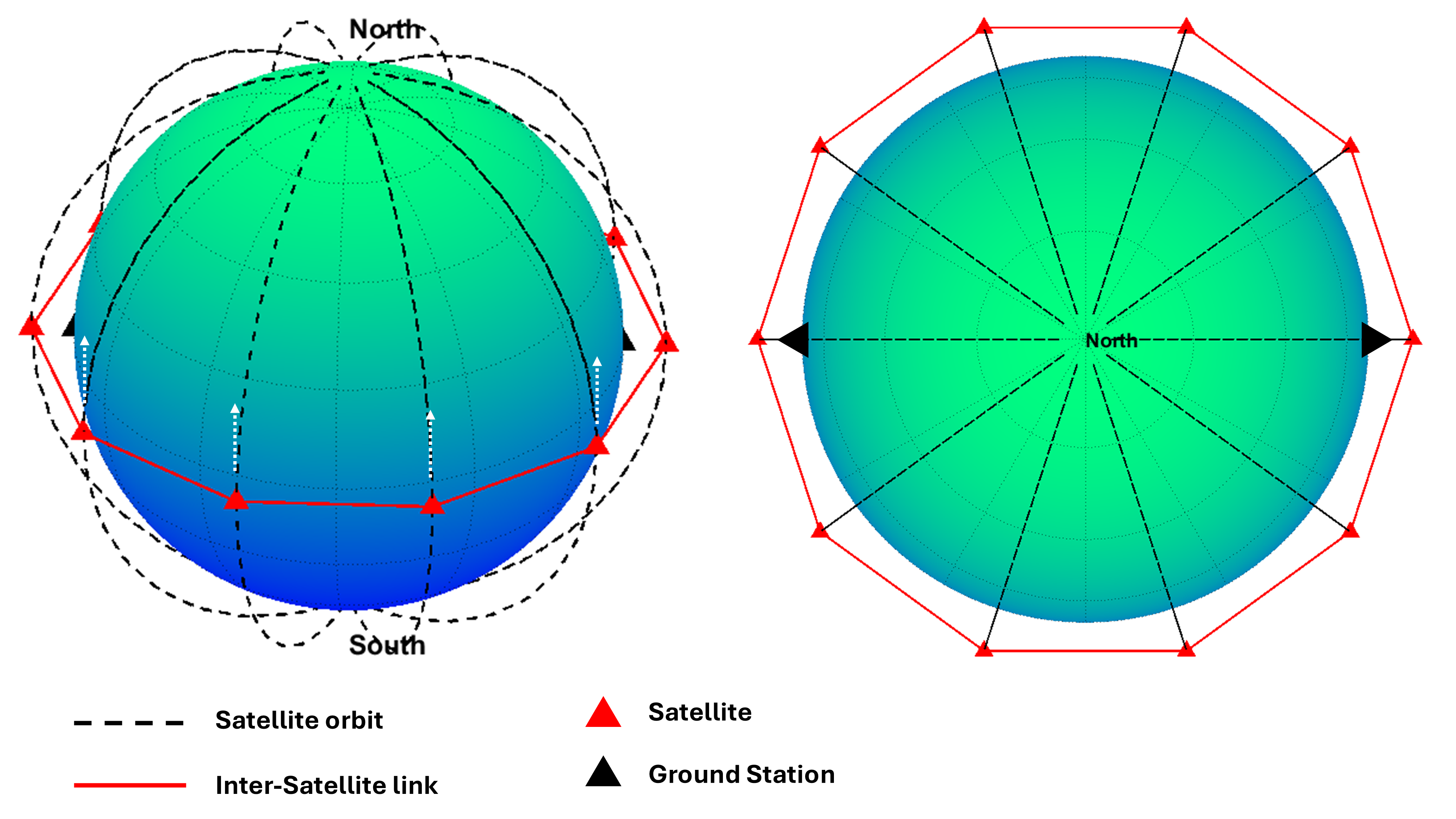}
\caption{Illustration of the Type-1 constellation, consisting of equally spaced polar orbital planes with one satellite per plane. This geometry ensures continuous inter-satellite line-of-sight, thereby maintaining an unbroken ring network at all times. In this configuration, $GS_1$ and $GS_2$ are assumed to lie at the same latitude but on opposite sides of the Earth (a longitudinal separation of $180^\circ$).}

\label{fig_constellation_type_1}
\end{figure}

\textbf{Type-1 Constellation (Pulsating Polar ring)}:
In this configuration, satellites are placed in near-polar orbits with equal longitudinal spacing. Each orbital plane contains a single satellite, and all satellites share the same altitude. At any given time, the satellites lie approximately at the same latitude, forming a ring around the Earth. A sufficiently large number of satellites $(N_s>9)$ ensures satisfaction of the inter-satellite \gls{los} condition. This polar ring provides global coverage, allowing every ground station to be overflown at least once per orbital period. For a ground station located at mid-latitudes, a zenith pass (zenith angle $0°$) yields a maximum ground–satellite \gls{qkd} session of approximately 294 seconds per orbit, with all non-zenith passes having shorter durations and higher losses. \\

\begin{figure}[htbp]
\centering
\includegraphics[width=1.0\columnwidth]{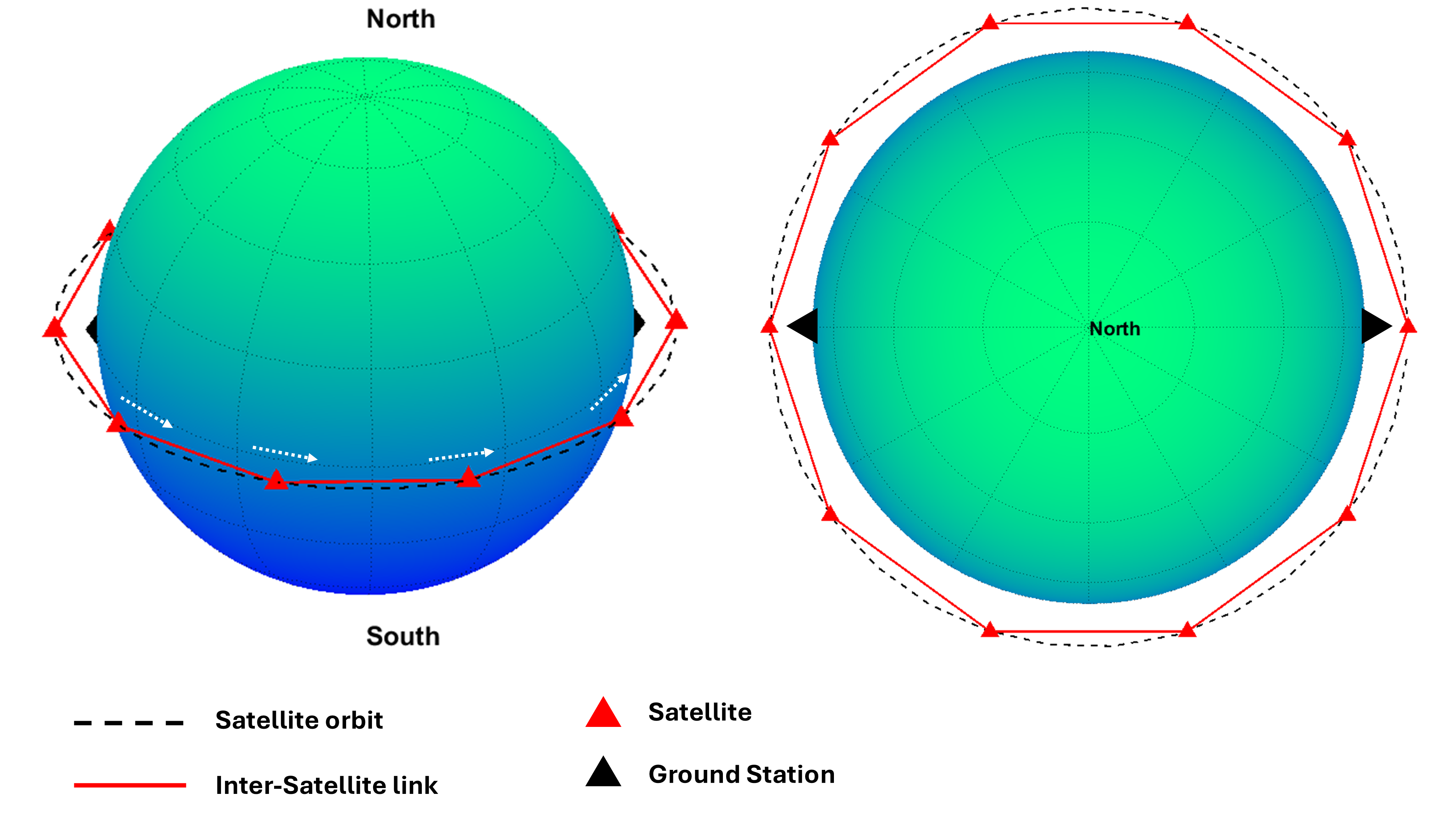}
\caption{Illustration of the Type-2 constellation, consisting of an equatorial orbit with all satellites evenly placed in the orbit. This geometry ensures continuous inter-satellite \gls{los} and offers a high visibility ratio. In this configuration, $GS_1$ and $GS_2$ are assumed to lie at the same latitude but on opposite sides of the Earth (a longitudinal separation of $180^\circ$).}

\label{fig_constellation_type_2}
\end{figure}

\textbf{Type-2 Constellation (Equatorial Ring)}:
The second configuration consists of satellites in equatorial circular orbits with uniform angular spacing, also satisfying the \gls{los} condition for $(N_s>9)$. Although this constellation does not provide global coverage, it offers an extended visibility ratio $\rho_{\text{vis
}}$ for ground stations located near the equator. For sufficiently large constellation sizes $(N_s\geq20)$, the equatorial ring enables near-continuous \gls{qkd} connectivity for equatorial ground stations, providing operational characteristics similar to terrestrial fiber-based \gls{qkd} networks.

\section{Redundant End-to-End Key Exchange over the Satellite Ring}
\label{sec_e2e_key_forward}

The redundant key-exchange protocol introduced in~\cite{calsi2024end} employs a sequence of \gls{tfqkd} links to relay key material across the network, enabling end-to-end key establishment. In our model, the ground stations act as end users, while satellites serve as intermediate nodes, such that secrecy is established directly between the ground stations. Because each hop forwards keys via \gls{tfqkd} rather than regenerating them, an adversary must compromise at least two consecutive nodes along a path to gain useful information. When multiple node-disjoint paths are available, the attacker must breach adjacent nodes on every path, substantially enlarging the attack surface. This stands in contrast to trusted-node architectures, where compromising a single intermediary can reveal the entire key. The proposed redundancy therefore strengthens security, minimizes reliance on localized trust, and improves network robustness.

All classical channels used for \gls{qkd} coordination are assumed to be authenticated, consistent with standard security frameworks. Node identities may be verified using conventional or information-theoretically secure authentication schemes. In the \gls{leo} ring topology, adjacency relations are determined by predictable orbital dynamics (e.g., \gls{los} conditions) and maintained through authenticated control signaling.

\subsection{Key Forwarding Mechanism}

The redundant key-forwarding protocol is implemented over the satellite \gls{qkd} ring to enable end-to-end key establishment between distant ground stations. When Alice ($GS_1$) and Bob ($GS_2$) are connected to satellites $S_i$ and $S_k$, respectively, the ring is partitioned into two directed segments,

$\Gamma^{+} : 
GS_1 \rightarrow S_i \rightarrow S_{i+1} \rightarrow \cdots \rightarrow S_k \rightarrow GS_2$,

$\Gamma^{-} : 
GS_1 \rightarrow S_i \rightarrow S_{i-1} \rightarrow \cdots \rightarrow S_k \rightarrow GS_2$,

with indices taken modulo $N_s$. Alice generates two independent keys, $X^+$ and $X^-$, forwarding one along each segment to provide redundancy.

We describe the procedure along $\Gamma^{+}$; $\Gamma^{-}$ follows symmetrically.

\paragraph*{Initialization and link-key generation}

Alice performs \gls{tfqkd} with $S_{i+1}$ to generate $K_{i-1,i+1}$ and establishes a point-to-point \gls{qkd} key $\bar{K}_{i-1,i}$ with $S_i$.  
Each intermediate satellite $S_j$ generates a \gls{tfqkd} key $K_{j,j+2}$ with $S_{j+2}$, continuing sequentially until $S_{k-2}$ establishes $K_{k-2,k}$.  
Bob performs \gls{tfqkd} with $S_{k-1}$ to generate $K_{k+1,k-1}$ and shares a point-to-point key $\bar{K}_{k,k+1}$ with $S_k$.  

Consequently, every node along the segment possesses two local quantum keys.

\paragraph*{XOR-based forwarding}

Alice samples an $n$-bit secret $X^+\in\{0,1\}^n$ and computes
$
X_{i-1}
=
X^+\oplus \bar{K}_{i-1,i}\oplus K_{i-1,i+1},
$
which is forwarded to $S_i$.

At the first satellite hop ($S_i$), the value is updated using the locally available keys, \\
$
X_i
=
X_{i-1}\oplus \bar{K}_{i-1,i}\oplus K_{i,i+2}
=
X^+\oplus K_{i-1,i+1}\oplus K_{i,i+2},
$
and transmitted to $S_{i+1}$.

For intermediate satellites $S_j$ ($i<j<k$), forwarding involves only inter-satellite \gls{tfqkd} keys: \\
$
X_j
=
X_{j-1}\oplus K_{j-1,j+1}\oplus K_{j,j+2}
=
X^+\oplus K_{j-1,j+1}\oplus K_{j,j+2},
$
which is passed to $S_{j+1}$.

At the final hop, $S_k$ computes \\
$
X_k
=
X_{k-1}\oplus K_{k-2,k}\oplus \bar{K}_{k,k+1}
=
X^+\oplus K_{k-2,k}\oplus \bar{K}_{k,k+1},
$
and forwards it to Bob. Bob then recovers \\
$X^+
=
X_k\oplus \bar{K}_{k,k+1}\oplus K_{k+1,k-1}$.

Executing the same procedure independently along $\Gamma^{-}$ yields $X^-$.  
The final end-to-end secret key is therefore
$X_{\text{ring}} = X^+ \oplus X^-$, providing redundancy and increasing the adversarial effort required for compromise.

\subsection{Improved Security}

The proposed protocol combines quantum-generated link keys with XOR-based aggregation to realize end-to-end secrecy over the satellite ring. At each hop, the forwarded value is masked by two independent quantum keys, ensuring that neither intermediate satellites nor external adversaries ever access $X^{+}$ or $X^{-}$ in the clear. Unlike conventional \gls{tn} architectures—where compromising a single relay can expose the full key—the present design eliminates single-point failure and provides strictly stronger security guarantees.

Compromising a key propagated along one directed segment ($\Gamma^{+}$ or $\Gamma^{-}$) requires obtaining all masking keys protecting the forwarded value. Owing to the XOR structure, this necessitates compromising at least \emph{two consecutive} satellites on that segment. Since the final key is derived from two independent, node-disjoint segments, both must be broken to recover the end-to-end secret. The minimum number of compromised satellites depends on whether the attachment nodes $S_i$ and $S_k$ are affected:

\begin{itemize}
    \item \textbf{If $S_i$ or $S_k$ is compromised:}  
    Because these nodes belong to both segments, at least \emph{three} compromised satellites are required.

    \item \textbf{If neither $S_i$ nor $S_k$ is compromised:}  
    The segments are internally disjoint, and the adversary must compromise two consecutive satellites on each segment, for a minimum of \emph{four} satellites.
\end{itemize}

Hence, compared to trusted-node networks, the protocol substantially enlarges the adversarial attack surface. While a single compromised node may disrupt availability (e.g., via denial-of-service), secrecy remains intact unless two consecutive satellites within a segment are breached. Additional independent segments or multiple rings can further enhance resilience. A general security analysis is provided in Sec.~\ref{sec:sec-analysis}.

\section{Network implementation}
\label{sec_network_imp}

\subsection{Link Budget}
\label{sec:qkd-links}

\subsubsection{Uplink}
\label{subsec:uplink}

The GS$\rightarrow$Sat uplink is primarily limited by atmospheric turbulence, extinction, pointing errors, and zenith-angle–dependent path elongation. Following the turbulence-aware model of~\cite{behera2024estimating}, the instantaneous channel efficiency is
\begin{equation}
    \eta_{UL}
    = \eta_{opt} \, \eta_{atm}^{\sec(\zeta)} \,
      L_{fs} \, G_t \, G_r \, \langle \eta_{turb} \rangle ,
    \label{eq:ul_power_compact}
\end{equation}
where $\eta_{opt}$ denotes intrinsic optical losses, $\eta_{atm}^{\sec(\zeta)}$ represents elevation-dependent extinction, and $L_{fs}=(\lambda/(4\pi L))^2$ is the free-space loss. The transmitter gain $G_t=8/\Theta_B^2$ depends on the beam divergence $\Theta_B$, while the receiver gain $G_r=4\pi A_r/\lambda^2$ is determined by the satellite aperture $A_r$.

Atmospheric turbulence is characterized by the refractive-index structure parameter $C_n^2(h)$ (Hufnagel–Valley model), from which the coherence length $r_0$ is obtained via path integration scaled by $\sec(\zeta)$. The combined effects of beam spreading, beam wander, and scintillation are incorporated into the multiplicative factor $\langle \eta_{turb} \rangle$, which depends on $r_0$, receiver aperture $D_{rx}$, and wavelength, with detailed expressions given in~\cite{behera2024estimating}.

\begin{figure}[htbp]
\centering
\includegraphics[width=0.90\columnwidth]{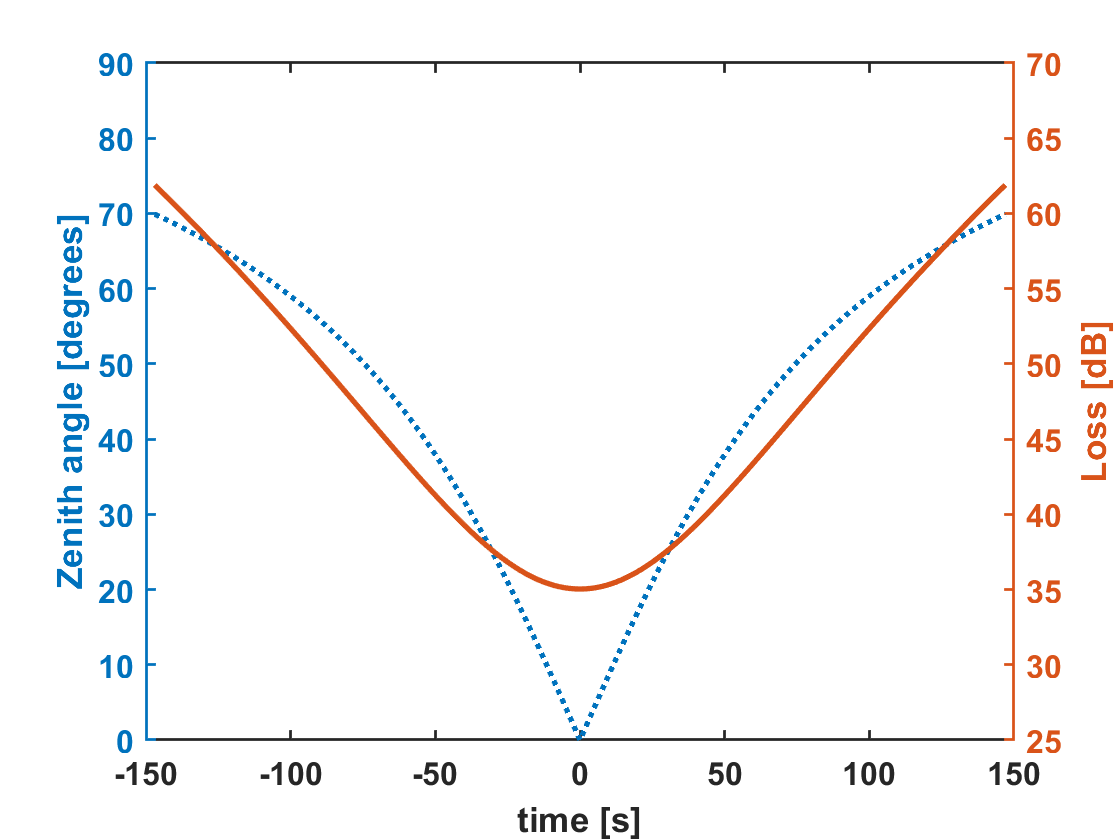}
\caption{Uplink channel loss as a function of the satellite pass. For the
ground-to-satellite uplink, a cutoff zenith angle of 70◦ is imposed. Under
this constraint, a zenith pass ( $\theta = 0^\circ$) yields a maximum ground–satellite
QKD session duration of 294 seconds.}
\label{fig_loss_UL}
\end{figure}

Figure~\ref{fig_loss_UL} shows the resulting uplink loss over a satellite pass. With a $70^\circ$ cutoff in zenith angle, a zenith pass provides a $294\,\mathrm{s}$ \gls{qkd} window, during which $\eta_{UL}$ is dominated by near-ground turbulence and rapidly increasing atmospheric airmass away from zenith.

\subsubsection{\gls{isl}}
\label{subsec:isl_budget}

The \gls{isl} operates in vacuum, so link behaviour is fully determined by the terminal optical transmission, the Gaussian-beam spreading over the instantaneous separation \(L\), and residual pointing jitter between satellites. The total efficiency is expressed as $\eta_{\mathrm{ISL}} = \eta_{\mathrm{opt}} \, \eta_{\mathrm{geo}}(L) \, \eta_{\mathrm{point}} $.

The optical term \(\eta_{\mathrm{opt}}\) captures all static transmission effects internal to the terminals. The geometric term \(\eta_{\mathrm{geo}}(L)\) depends on the diffraction-broadened Gaussian beam radius at distance \(w(L)\), computed from the beam divergence $\Theta_B$, wavelength $\lambda$, and resulting Rayleigh range $z_R$, while the receiver aperture \(D_{rx}\) determines how much of that beam is collected: $\eta_{\mathrm{geo}}(L) = 1 - \exp\!\left[ -2 \left( {D_{rx}/}{2\ w(L)} \right)^2 \right]$.

Pointing losses are driven by the effective optical gains of the transmitter and receiver, respectively set by the $\Theta_B$ and by the diffraction-limited aperture with the platform jitter variance \(\sigma_p^2\). The resulting pointing term is  $\eta_{\mathrm{point}} = \exp\!\left( -G_t \sigma_p^{2} \right)\, \exp\!\left( -G_r \sigma_p^{2} \right)$.

\begin{figure}[htbp]
\centering
\includegraphics[width=0.90\columnwidth]{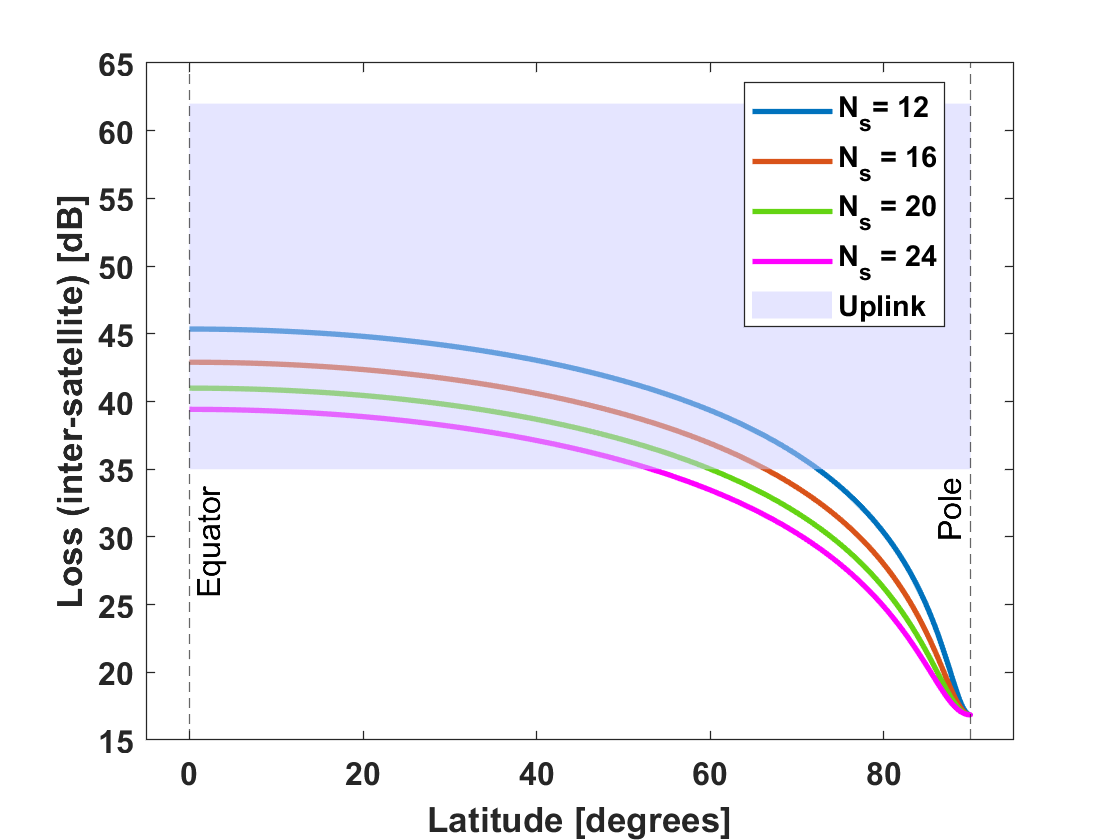}
\caption{\gls{isl} loss between $S_i$ and $S_{i+1}$ for the Type-I (polar) constellation. Increasing the number of satellites reduces the inter-satellite spacing, thereby lowering the optical loss. The shaded region indicates the range of uplink losses observed during a ground–satellite session. For comparison, the losses at the equator correspond to the Type-II (equatorial) constellation.}
\label{fig_loss_ISL}
\end{figure}

\subsection{QKD in the Network}
The key-forwarding protocol requires (i) point-to-point \gls{qkd} links between directly connected nodes and (ii) intermediate-node links for \gls{tfqkd}. Point-to-point links use decoy-state BB84, while intermediate-node links employ \gls{tfqkd} for long-distance operation beyond the repeaterless bound. In the satellite network, GS–Sat uplinks dominate performance due to diffraction, turbulence, and beam wander, with effective link efficiency $\eta_{j,i+1} = \max\{\eta_{UL}, \eta_{ISL}\}$ ranging from $\sim 70$–$144\,$dB as shown in Fig.~\ref{fig_loss_UL} and Fig.~\ref{fig_loss_ISL} and visibility-limited sessions of duration $\Delta t$.

The \gls{sns} protocol \cite{jiang2019unconditional} partitions each attempt into Z windows (probability $p_Z$) and decoy X windows ($1-p_Z$). In Z windows, weak coherent or vacuum states are sent probabilistically; in X windows, decoy intensities $(0, \mu_1, \mu_2)$ are used to estimate single-photon yields and phase-flip error $e_1^{ph}$ via phase-slice postselection ($\Delta$). Finite-key analysis over $\Delta t$ gives the secret-key length
\begin{equation}
\text{SKL} = n_1 (1-h(e_1^{ph})) - \lambda_{EC} - 2\log_2\frac{2}{\varepsilon_{\rm cor}}\frac{2}{\sqrt{2}\varepsilon_{\rm PA}\hat{\varepsilon}},
\end{equation}
where $n_1$ is the number of untagged single-photon bits, $\lambda_{EC} = f_{\rm EC}\,\bar{n}\,h(E_Z)$, and the failure parameters $\varepsilon_{\rm cor}, \varepsilon_{\rm PA}, \hat{\varepsilon}, \bar{\varepsilon}, \varepsilon_{n_1}$ combine into $\varepsilon_{\rm sec}$, with total $\varepsilon_{\rm tol} = \varepsilon_{\rm cor} + \varepsilon_{\rm sec}$.  

Over the total observation period $T_{\rm total}$, raw key bits are accumulated for all $\{GS_j, S_{i\pm1}\}$, optimizing the SNS parameters $\{\mu, \mu_1, \mu_2, p, p_Z, p_0, p_1, \Delta\}$ per link. The overall secret key generated is
\[
\mathrm{SKL}^{(T_{\rm total})}_{\rm protocol} = \sum_i \min\!\left\{ \mathrm{SKL}^{(T_{\rm total})}_{i\pm1,1},\, \mathrm{SKL}^{(T_{\rm total})}_{k\pm1,2} \right\},
\]
with the minimum enforcing the weaker GS–Sat segment as the bottleneck and $k = i + N_s/2$ for $GS_1$ and $GS_2$ separated by $180^\circ$ longitude.

\section{Results}
\label{sec_results}

\begin{table}[t]
\centering
\caption{Parameters used in the calculations.}
\renewcommand{\arraystretch}{1.1}
\setlength{\tabcolsep}{4pt}
\begin{tabular}{l c c}
\hline
\textbf{Parameter} & \textbf{Symbol} & \textbf{Value} \\
\hline
Earth radius & $R_E$ & 6371 km \\
Satellite altitude & $h$ & 500 km \\
Max. zenith angle & $\theta_{\max}$ & $70^\circ$ \\
Wavelength & $\lambda$ & 850 nm \cite{behera2024estimating} \\
Beam divergence & $\theta_{\mathrm{div}}$ & $15\,\mu\text{rad}$ \\
GS beam waist & $w_0$ & 27 cm \\
Atmospheric loss coeff. & $\alpha_{\text{atm}}$ & 1.55 dB \\
Pointing jitter & $\sigma_{\mathrm{jitter}}$ & $1\,\mu\text{rad}$ \cite{Milasevicius2024Review} \\
GS Transmitter diameter & $D_{tx}^{GS}$ & 54 cm \\
Sat. Transmitter diameter & $D_{tx}$  & 30 cm \\
Sat. Receiver diameter & $D_{rx}$ & 30 cm \\
Detector efficiency & $\eta_{\mathrm{det}}$ & 0.50 \cite{Ma2025Drone} \\
Dark count probability & $P_{\mathrm{dc}}$ & $10^{-9}$ \\
Optical error & $e_{\mathrm{opt}}$ & 0.05 \\
Error Correction inefficiency & $f_{EC}$ & 1.11 \\
QKD Repetition rate & $f_{rep}$ & 1 GHz \\
Security parameters & $\varepsilon_{cor}, \varepsilon_{PA}$ $\hat{\varepsilon}$. $\bar{\varepsilon}$, $\varepsilon_{n_1}$ & $10^{-10}$ \\
\hline
\end{tabular}
\label{tab:params}
\end{table}

\begin{figure*}
\centerline{\includegraphics[width=39pc]{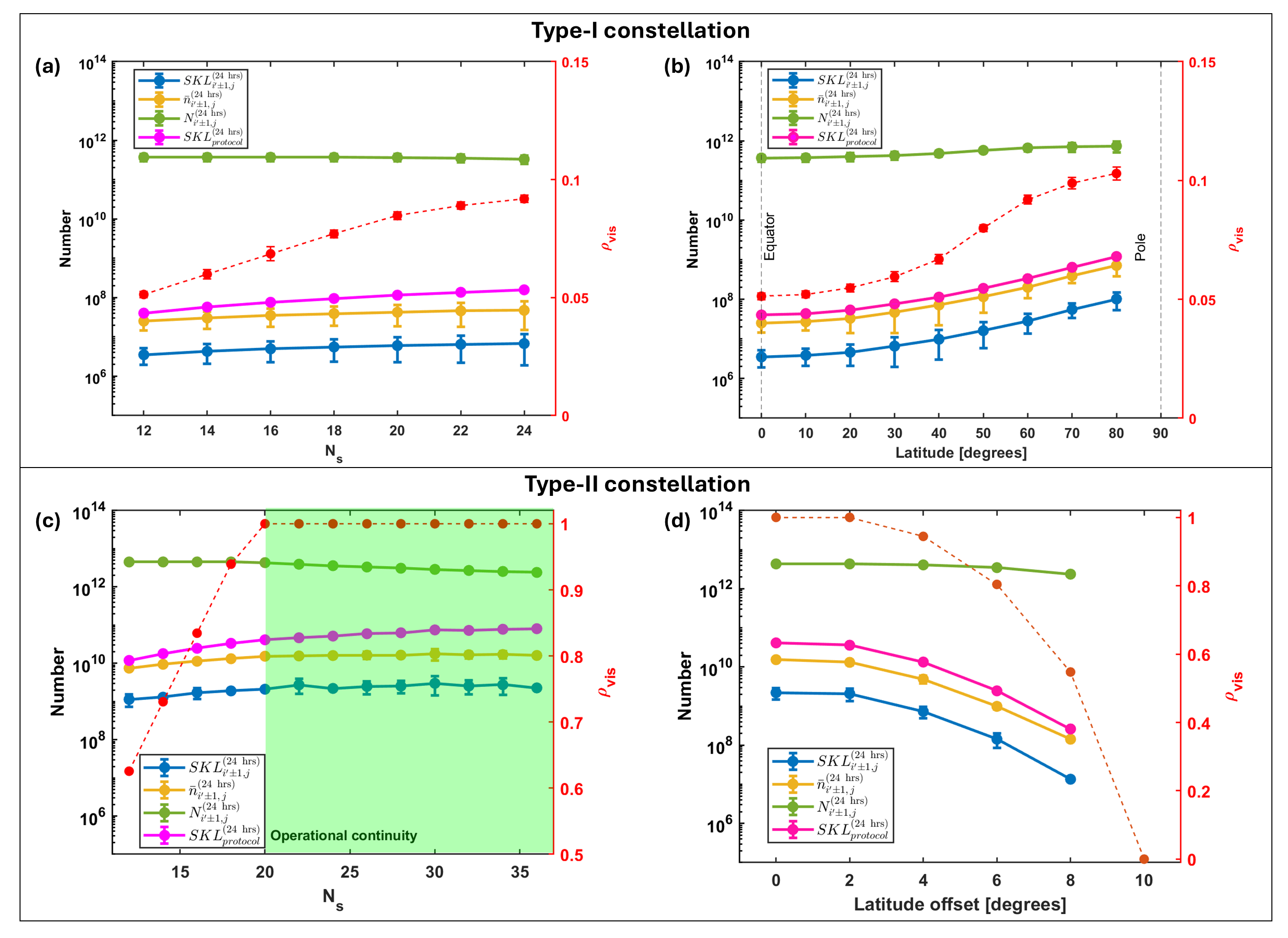}}
\caption{Daily key-generation statistics for the satellite–ring \gls{qkd} network. $GS_1$ and $GS_2$ share the same latitude and are separated by $180^\circ$ in longitude. The blue curve shows the per-satellite ground–satellite \gls{skl} over 24 hours; the orange curve gives the corresponding sifted-key bits. The magenta curve denotes the end-to-end \gls{skl} of the protocol, scaling approximately as $N_s \times$ (ground–satellite \gls{skl}). The green curve represents the daily block size (total \gls{tfqkd} signal states) used for finite-size estimation. Results are averaged over 30 simulated days with standard-deviation error bars.
\textbf{Type I constellation:} 
\textbf{(a)} Performance versus $N_s$ with $GS_j$ at the Equator. 
\textbf{(b)} Performance versus ground-station latitude for $N_s=12$.
\textbf{Type II constellation:} 
\textbf{(c)} Both the protocol key rate $\mathrm{SKL}^{(24\,\mathrm{hrs})}_{\mathrm{protocol}}$ and the satellite–ground $\mathrm{SKL}^{(24\,\mathrm{hrs})}_{i'\pm1,j}$ increase with $N_s$; operational continuity ($\rho_{\text{vis}}=1$) is achieved at $N_s=20$. 
\textbf{(d)} For $N_s=20$, shifting the ground-station latitude from the Equator reduces performance: $\mathrm{SKL}^{(24\,\mathrm{hrs})}_{\mathrm{protocol}}$ decreases sharply and vanishes at a $10^\circ$ offset, where $\rho_{\text{vis}}$ drops from 1 to 0.}
\label{fig_resesults}
\end{figure*}

We evaluate the performance of the proposed protocol for both Type~I and Type~II satellite constellations. Throughout the analysis, we assume that $GS_1$ and $GS_2$ are located at the same latitude with a longitudinal separation of $180^\circ$. The Type~I ring structure provides global coverage, as the ring continuously contracts and expands while sweeping from the northern to the southern hemisphere. Using the uplink and \gls{isl} loss characteristics shown in Fig.~\ref{fig_loss_UL} and Fig.~\ref{fig_loss_ISL}, we compute the secret-key length accumulated over a $24$-hour period, $\mathrm{SKL}^{(24\ \mathrm{hrs})}_{\mathrm{protocol}}$. All the simulation parameters are mentioned in Table~\ref {tab:params}.

The resulting performance is presented in Fig. \ref{fig_resesults}. As the ground-station latitude increases from the equator toward the poles, the values of $\mathrm{SKL}^{(24\ \mathrm{hrs})}_{i\pm1,\, j}$ improve due to higher visibility fractions $\rho_{\mathrm{vis}}$ and reduced ISL losses at higher latitudes. For a given constellation size, the total end-to-end key yield scales approximately as
$\mathrm{SKL}^{(24\ \mathrm{hrs})}_{\mathrm{protocol}}
    \approx N_s \times \mathrm{SKL}^{(24\ \mathrm{hrs})}_{i\pm1,\, j}$.

We also show the corresponding block size (i.e., the total number of signal states sent in the \gls{tfqkd}) $N^{(\mathrm{Total})}_{i'\pm1,\, j}$ and raw-key length $\bar{n}^{(\mathrm{Total})}_{i'\pm1,\, j}$ accumulated over the 24-hour period, providing a detailed overview of the \gls{tfqkd} performance within the satellite network. For the Type~I constellation, the protocol generates $\mathrm{SKL}^{(24\ \mathrm{hrs})}_{\mathrm{protocol}}$ $\approx40.2$ Mbits/day and $\approx 52.8$ Mbits/day at the Equator and the Poles, respectively, with $N_s = 12$. Whereas on increasing $N_s$ from 12 to 24 at Equator, we find $\mathrm{SKL}^{(24\ \mathrm{hrs})}_{\mathrm{protocol}}$ significantly rising from $\approx40.2$ Mbits/day to $\approx 165.6$ Mbits/day.

We evaluated the performance of the Type~II constellation, with the results shown in Fig.~\ref{fig_resesults}. For a ring size of $N_s = 12$, the protocol achieves a \gls{skl} of $11.87~\mathrm{Gbits/day}$. Increasing the number of satellites improves the visibility ratio, which reaches unity at $N_s = 24$. At this point, each ground station $GS_j$ maintains continuous connectivity to the satellite ring, enabling uninterrupted operation analogous to terrestrial \gls{qkd} networks. The corresponding protocol \gls{skl} increases to $41.18~\mathrm{Gbits/day}$, and continues to improve with larger constellation sizes. For example, with $N_s = 36$, the \gls{skl} reaches $80.61~\mathrm{Gbits/day}$. Although the Type 2 network is optimum for \glspl{gs} on Equator, it can still work if the latitude of a \gls{gs} is offset from Equator by an angle below $10^\circ$. At a $10^\circ$ offset, $\rho_{\text{vis}}$ drops to 0.

These results clearly demonstrate that the network performance scales favorably with the constellation size. Moreover, the security threshold also goes up with constellation size, which is explained in detail in \ref{sec:sec-analysis}

\subsection{Generalized Network and Security Analysis}
\label{sec:sec-analysis}

Our network follows the non-localized trust model of~\cite{calsi2024end}, providing resilience against intermediate-node compromise. In the redundant ring architecture, an adversary must compromise at least \emph{two consecutive satellites} on each segment ($\Gamma^{+}$ and $\Gamma^{-}$) to recover the secret key, i.e., a minimum of four nodes. If one attachment node ($S_{i'}$ or $S_{k'}$) is already compromised, this reduces to three.

Security can be further enhanced in two ways. First, by adding independent rings, an adversary must compromise three consecutive satellites \emph{per ring}, enlarging the attack surface. Second, by extending \gls{tfqkd} connectivity beyond nearest-neighbour links $(i,i+2)$ to $(i,i+2),\dots,(i,i+r)$, each node and ground station can XOR a larger set of independent keys. Under this generalized scheme, compromising a segment requires $r$ consecutive nodes, or $2r-1$ in the worst case if an attachment node is compromised.

Adding rings is straightforward, as each ring operates independently. Extending \gls{tfqkd} to the $r^{\rm th}$ neighbour requires line-of-sight and manageable \gls{isl} losses, which improve as the number of satellites $N_s$ increases. For ISL losses below $45~\mathrm{dB}$, $r=3$ is achievable for $N_s>24$, $r=4$ for $N_s>36$, and $r=5$ for $N_s>48$. Hence, increasing $N_s$ simultaneously strengthens security and enhances visibility fraction $\rho_{\rm vis}$, boosting secret-key rates.

In a generalized architecture with $N_R$ independent rings and up to the $r^{\rm th}$ neighbouring \gls{tfqkd} links, the network remains secure as long as fewer than $2r-1$ consecutive satellites in a ring or $r$ consecutive per segment are compromised. In the worst-case scenario, an adversary must compromise at least $N_R(2r-1)$ satellites to break the protocol, demonstrating how both security and performance scale favorably with constellation size.

\section{Discussion and Conclusion}
\label{sec_discussion}

We proposed a ring-based satellite \gls{qkd} network with two constellation types: Type 1 provides global coverage, while Type 2 ensures operational continuity near the Equator. The network employs a key-forwarding mechanism where the end-user key is XOR-ed with multiple \gls{qkd} keys at each hop, ensuring that the secret is never exposed to intermediate satellites. To recover the key, an adversary must compromise at least $r$ consecutive nodes per segment when \gls{tfqkd} is extended up to the $r^{\rm th}$ nearest neighbour, providing significantly stronger security than conventional trusted-node schemes. Adding more independent rings and increasing $r$ further enlarges the adversarial attack surface. Network performance scales favorably with constellation size: both security and secret-key rates improve as the number of satellites $N_s$ increases. Our \gls{skl} results demonstrate that key rates rise with $N_s$, and Type 2 constellations achieve continuous operation once $N_s \ge 20$.  

Recent technological advances—including high-speed satellite-compatible QKD~\cite{roger2023real}, high-loss-tolerant TF-QKD~\cite{wang2022twin}, and integrated microsatellite–portable-ground-station platforms~\cite{li2025microsatellite}—support the feasibility of the proposed architecture. While specialized quantum hardware is not required, practical implementation remains challenging due to phase stabilization over long distances, precise alignment in dynamic orbits, and establishing quantum uplinks. Large-scale deployment also entails substantial financial investment. Addressing these technical and economic challenges will be essential for experimental validation and operational deployment of the network.



\bibliography{references}
\bibliographystyle{IEEEtran}

\end{document}